\documentclass{emulateapj}

\usepackage{hyperref}
\usepackage{graphicx}

\begin{document}


\title{The Extended Tidal Tails of NGC 7089 (M2)}

\author{Carl J. Grillmair}
\affil{IPAC, California Institute of Technology, Pasadena, CA 91125}
\email{carl@ipac.caltech.edu}

\begin{abstract}

  Using photometry and proper motions from Gaia Early Data Release 3,
  we detect a $45\arcdeg$-long trailing stellar debris stream
  associated with the old, metal-poor globular cluster NGC 7089. With
  a width on the order of 100 pc, the extended stream appears to be as
  dynamically cold as the coldest known streams found to date.  There
  is some evidence for an extended leading tail extending between
  $28\arcdeg$ and $37\arcdeg$ from the cluster, though the greater
  distance of this tail, combined with proper motions that are
  virtually indistinguishable from those of foreground stars, make the
  detection much less certain. The proper motion profile and the path on the
  sky of the trailing tail are not well matched using a simple
  Galactic potential comprised purely of a disk, bulge, and spherical
  halo. However, the addition of a moving, massive (M
  $= 1.88 \times 10^{11} M_\sun$) Large Magellanic Cloud
  brings the model predictions into much better agreement with the
  observables. We provide tables of the most highly ranked candidate
  stream stars for follow-up by ongoing and future spectroscopic
  surveys.

\end{abstract}



\section{Introduction}

The inner Galactic halo is now known to be populated with dozens of
stellar debris streams \citep{grillmair2016, shipp2018, malhan2018b,
  ibata2019, palau2019, ibata2021}. Many of these streams are relatively
narrow, with physical widths on the order of 100 pc, and they are
consequently assumed to have been produced by globular clusters. 

With episodic stripping and consequent variations in stream density,
tidal tails extending from extant globular clusters may be somewhat
problematic for detecting signatures of dark matter subhalos
\citep{kupper2012}. On the other hand, by providing well-characterized
progenitors, such tidal tails will be useful for understanding the
detailed physics of tidal stripping \citep{balbinot2017}, the
accretion sequence of the halo, and the shape of the Galactic
potential \citep{bovy2016, price-whelan2014, malhan2019,
  garavito-camargo2021}. Longer streams and streams on more eccentric
orbits are particularly sensitive to the shape of the halo potential
\citep{bonaca2018}.  To extract as much information as possible, it
would therefore be desirable to trace known streams for as far along
their trajectories as possible.

With the third data release (EDR3) of the Gaia catalog \citep{gaia2021} we now have
considerably more information for sorting stars into
substructures. \citet{malhan2018a}, \citet{malhan2018b},
\citet{ibata2019}, and \citet{ibata2021} have applied their
STREAMFINDER orbital integration technique to dozens of halo streams
that would have been impossible to detect with the purely photometric
techniques of a few years ago. Here we use an alternative, more
directed method to increase the known extent of the tidal tails of the
globular cluster NGC 7089. The extended envelope and incipient tidal
tails of NGC 7089 were first detected by \citet{grillmair1995} and
were among the strongest detections in their sample. \citet{kuzma2016}
subsequently extended these results to nearly two degrees, several
times the nominal tidal radius of the cluster. Most recently, tidal
tails extending between $8\arcdeg$ and $13\arcdeg$ were detected by
\citet{ibata2021} using Gaia EDR3 and STREAMFINDER.

Section \ref{analysis} describes our method, which largely mirrors
that used by \citet{grillmair2019} to detect the extended trailing
tail of M5. We discuss the apparent trajectories of the tails in
Section \ref{discussion}, where we also provide tables of our highest
ranked stream star candidates.  We make concluding remarks in Section
\ref{conclusions}.

\section{Analysis} \label{analysis}

Our analysis closely follows that of \citet{grillmair2019} as applied
to the globular cluster M5, combining color-magnitude and proper
motion filtering with orbit integrations and predictions based on
modeling the stripping of stars from NGC 7089. Here we made use of
Gaia EDR3 \citep{gaia2021}, whose proper motion
uncertainties are reduced by a factor of $\sim 2$ compared with Gaia
Data Release 2. We computed the expected proper motions and
trajectories on the sky using the Galactic model of \citet{allen1991},
coded in the IDL language and updated using the parameters of \citet{irrgang2013}. While this model
is fairly simple, assuming a $6.6 \times 10^{10} M_\sun$ Miyamoto-Nagai disk \citep{miyamoto1975},
a $9.5 \times 10^9 M_\sun$ spherical bulge, and a $1.8 \times 10^{12}
M_\sun$ spherical halo, our experience using it with other clusters
(e.g. \citet{grillmair2019}) suggests that it provides reasonable
approximations for actual cluster orbits in the inner halo. With
the updated parameters the model predicts M(R$ < 50$ kpc) $= 5.1 \times
10^{11} M_\sun$, which is slightly more than the M(R$ < 50$ kpc)$ = 4.6
\times 10^{11} M_\sun$ predicted by the \citet{wilkinson1999} model,
though considerably less than the M(R$ < 50$ kpc)$ = 8.1 \times 10^{11}
M_\sun$ predicted by the \citet{navarro1997} model.

Recent investigations \citep{kallivayalil2013, erkal2019, shipp2021}
suggest that the Large Magellanic Cloud (LMC) is considerably more
massive than previously supposed and may be a significant perturber of
satellite orbits. As an experiment we consequently augmented
our Galactic potential with a moving LMC, modeled as a point mass with
$M =1.88 \times 10^{11} M_\sun$ \citep{shipp2021}, arriving at its
present location on a first pass. The LMC trajectory was modeled as a
fall from $\approx 700$ kpc, slowed by dynamical friction
\citep{chandrasekhar1943} according to the local density of stars and
dark matter along its orbit. While the stellar/dark matter wake of the
LMC is believed to be quite substantial \citep{conroy2021,
  garavito-camargo2021}, we have not attempted to model it as an
additional perturber on the orbit of NGC 7089.

For modeling the orbit of NGC 7089 and the motions of associated
stream stars, we used a cluster distance of 11.693 kpc
\citep{baumgardt2021}, cluster proper motions of 
$\mu_{\alpha} = 3.435 \pm 0.025$ mas/yr and 
$\mu_\delta = -2.159 \pm 0.024$ mas/yr \citep{vasiliev2021}, and a
radial (line-of-sight) velocity of -3.72 km sec$^{-1}$ \citep{vasiliev2019}.  We
computed the predicted trajectories of the tidal tails by modeling the
release of stars from the cluster over time \citep{kupper2015,
  bowden2015, dai2018}. Specifically, the model tails were generated
by releasing stars from the cluster at a rate proportional to $1/R^3$,
where $R$ is the Galactocentric radius of the cluster at any point in
time. Escaping stars were simply placed at the current $\approx 70$ pc
tidal radius of the cluster \citep{harris1996}, both at the L1 and L2
points, and then integrated independently along their individual
orbits. Orbit integrations were typically run for the equivalent of 4
Gyrs to provide tidal tails long enough to cover our field of
interest. Running such integrations for 10 Gyrs produced essentially
identical trajectories but generated many overlapping streams and
scattered stars that only served to complicate our analysis.

The results of our modeling are shown in Figure 1. Of particular note
are the effects on cluster orbits and stream models of a massive
LMC. It is also interesting that, without a perturbing LMC, the tidal
tails closely follow the orbit of the cluster itself.  On the
other hand, including a massive LMC produces significant departures
between the tails and the similarly integrated orbit of the
cluster. Additional models were run without a massive LMC and varying the cluster proper
motions, radial velocity, and distance over their respective
uncertainty ranges. However, the results departed only very slightly from
the unperturbed models shown in Figure 1. The differences between the
perturbed and unperturbed models dwarf the variations that could be
attributed to the reported uncertainties in the observables.

Also shown in Figure 1 are the stream stars detected by \citet{ibata2021} with
STREAMFINDER. These stars match the predictions of our modeling quite well but,
due to their limited extent from the cluster, do not appear to favor one model
over the other.  This underscores the value of streams that trace an appreciable
fraction of their orbits, and extending NGC 7089's tidal tails considerably in
both directions should provide us with another powerful and sensitive probe of
the Galactic potential.

To reduce contamination by foreground stars, we limited our analysis to
stars with $18.3 < G < 20.0$, which for NGC 7089 encompasses the upper
main sequence, the turn-off region, and the subgiant branch. While the
quality of the photometry improves significantly up the red giant
branch (RGB), the RGB colors overlap those of the much larger
population of foreground stars. Experiments that included the RGB in
our filtering yielded significantly noisier results. For stars fainter
than $G = 20$, the Gaia photometry and proper motion
uncertainties become large enough that we again saw noticeably noisier
results.

We used a modified form of the matched filter described by \citet{rockosi2002}
and \citet{grillmair2009}. Specifically, we weighted stars both by their
position in the extinction-corrected $G_\odot, (G_{BP} - G_{RP})_\odot$
color-magnitude diagram of NGC 7089 and by their departures from the predicted
proper motion profiles shown in Figure 1. The Gaia photometry was corrected for
extinction using the reddening maps of \citet{schlegel1998}, themselves
corrected using the prescription of \citet{schlafly2011}, and using the Gaia DR2
coefficients derived by \citet{gaia2018}. Since the proper motions at any given
R.A. are not unique, we additionally imposed the predicted trajectories of the
leading and trailing arms on the sky so that the proper motion filtering applied
to any particular star would depend on which arm it is closest to. We then
summed the resulting filter signals by sky position to produce the weighted
surface density maps shown in Figure 2.

\section{Discussion} \label{discussion}

Figure 2 shows a fairly convincing ``string of pearls'' extending $\approx
45\arcdeg$ to the southwest of NGC 7089 to [R.A., Dec] = [$281\arcdeg,
  -16.9\arcdeg$]. Using proper motion profiles based purely on the
\citet{allen1991} model, this trailing stream is visible, but somewhat muted
compared with the results we obtain when we include a moving, massive, LMC.
Using a range of LMC masses, from $1 \times 10^{11} M_\sun$ to $4 \times 10^{11}
M_\sun$ did not noticeably improve the resulting signal strength or length of
the stream.  As our goal is not to try to constrain the mass of the LMC but
simply to identify stars in the extended tidal tails of NGC 7089, we adopted
$M_{LMC} =1.88 \times 10^{11} M_\sun$ determined by \citet{shipp2021} for all
subsequent analyses.

The various portions of the trailing tail are visible only
over a limited range of assumed distance moduli, strengthening and
subsiding with increasing distance in the manner we would expect from
the photometric uncertainties.  The narrowness of the trailing tail
(FWHM $\approx 0.7\arcdeg = 110$ pc) indicates a dynamically cold
structure comparable with the Pal 5 and GD-1 streams
\citep{odenkirchen2001, odenkirchen2009, grillmair2006, malhan2019, gialluca2021}.
Though we attempted to trace the trailing tail across the Galactic
plane to the northern Galactic hemisphere, we were unable to find an
unambiguous continuation of the stream.

The western end of the trailing tail sits about $4\arcdeg$ south of
the predicted trajectory with no perturbing LMC. However, the addition
of a first-pass, massive LMC brings the trailing stream model into excellent
agreement with the observed tail. While the extended potential of the
Galaxy is now thought to be quite complicated and far from axisymmetric
\citep{garavito-camargo2021}, the relatively simple Galactic model of
\citet{allen1991}, combined with a massive, moving LMC, is evidently
sufficiently accurate for predictive work in the inner regions of the
Galaxy.

To the northeast of NGC 7089, the leading tail is considerably less
obvious. This is presumably because ($i$) the expected distance of
this portion of the stream ($\approx 20 $ kpc) is significantly
greater than that of NGC 7089, thereby substantially reducing the number
of stream stars brighter than our $G = 20$ limit, and ($ii$) the
predicted proper motions of stream stars become nearly zero and thus
very similar to those of the much larger population of field stars.

There is a tenuous cloud of signals at [R.A., dec] $\approx$
[$347\arcdeg, 3\arcdeg$] that corresponds roughly with the expected
apogalactic bend in NGC 7089's orbit computed either with or without a
massive LMC, as well as the stream model computed without a massive
LMC. However, there is also a very tenuous feature within the bounds
of the model stream trajectory computed with a moving, massive
LMC. This is more easily seen in Figure 3, where we show the filtered
image in Figure 2 overlaid with the trajectories shown in Figure
1. While our sampling of stream stars is limited by our magnitude
cut-off, we should still be benefitting from the expected orbital
convergence of stars at apogalacticon. Our stream models indicate that
the surface density of stars should be more than five times higher
near apogalacticon than it is over the observed length of the trailing
tail. Taking a typical globular cluster luminosity function, this
overdensity should effectively overcome the loss of the faintest
$\approx 1.1$ magnitudes of our sample.  However, the much increased
contamination by field stars with similar proper motions appears to
have effectively nullified this enhancement.

There is an interesting, relatively strong, triple-lobed signal at
[R.A., Dec]=[$346.75\arcdeg, -9.75\arcdeg$], lying along both the
cluster orbit and the model stream computed using a high-mass LMC,
that could conceivably be part of the leading tail. If we regard this
feature as a legitimate stream signal then the arc length of the
leading tail as measured from the cluster would be 
$37\arcdeg$ along the orbit of NGC 7089, and $28\arcdeg$ along the
stream model.

We see no signs of a leading stream extending significantly past
apogalacticon, along the infalling portion of the orbit. Though these
stars are expected to be at distances of between 6 and 10 kpc and
therefore well within our reach, it may be that their surface density
is simply too low to be apparent in Figure 2. These stars would be
accelerating towards perigalacticon, reducing their
surface density by a substantial factor. The stream models in Figure 1
show that in each case, this infalling portion of the orbit is
significantly depleted, with virtually no stars evident for the
LMC-perturbed stream model. While certain individual peaks in Figure 2 may
indeed belong to the NGC 7089's leading tail, their surface density is
very low and not noticeably different from that of the surrounding
field.

Our matched-filtering ensures that the stars with the strongest signals have
colors, magnitudes, and proper motions close our expectations. The agreement
between the trajectory of the trailing tail and the massive-LMC stream model is
reassuring, and supports the hypothesis that the LMC is both massive and
consequential for the motions of stars in the inner halo. However, Figures 2 and
3 by themselves do not readily convey the uncertainties remaining. Figure 4
compares the distances of our candidate stars, estimated both photometrically
and using Gaia EDR3 parallaxes, with the distances predicted by our stream
models. The photometric distances were estimated by shifting NGC 7089's
$G_\odot, (G_{BP} - G_{RP})_\odot$ locus brightwards and faintwards until the
colors and magnitudes of individual stream candidates matched that of the locus
at some point along its length. This would often occur at two points along the
locus, depending on whether the star was on the main sequence or the subgiant
branch. In these cases we would look to the EDR3 parallax measurements to
resolve the ambiguity. Since we want to compare distances for individual stars,
we estimated distance using $d = 1/\varpi$. In many cases the uncertainties in
the parallax measurements were too large to be informative, and in those cases
we simply chose the distance that best matched the predictions of our model.

Our estimated distances generally agree to within $1\sigma$ with both
model predictions. On the other hand the five western-most stars
appear to depart systematically from both the unperturbed and
  LMC-perturbed models, lying between 1 and 2 kpc (0.3 to 0.5 mag)
closer to us than the models would suggest. This may be due to
inadequacies in our assumed potential near the Galactic center (the
lack of a Galactic bar, for example), or it may be that these stars
are in a chance alignment and are not in fact part of the
stream. Radial velocity measurements will be required to resolve this
ambiguity.

A positional match of our highest-weighted stars with the Sloan
Digital Sky Survey Data Release 15 \citep{aguado2018} as well as the
LAMOST Data Release 6 \citep{zhao2012} did not yield any radial
velocity measurements outside the immediate environs of NGC 7089. To
aid in radial velocity follow-up, Tables 1 and 2 list the stars with
the highest matched-filter signals in each of the trailing and leading
tails, respectively. We applied a signal strength threshold to keep
the tables to a manageable size while still capturing the entire
extent of the trailing tail in Figure 2. The leading tail candidates
include both stars in the putative apogalactic cloud and in the more
southerly arm predicted using the LMC-perturbed stream
model. Identifying bona fide stream stars at or near apogalacticon is
especially important as they are particularly sensitive to the size
and shape of the halo potential \citep{bonaca2018}. The positions and
proper motions of all stars in Tables 1 and 2 are plotted in Figure
5. We note that of the 18 stars designated potential stream stars by
\citet{ibata2021}, only four are contained in Tables 1 and 2. Four of
their stars fall below our 18.3 magnitude cutoff and one falls
outside our region of interest (i.e. within $0.6\arcdeg$ of the
cluster itself). The remaining nine stars are in our larger sample,
but their relative signal strengths fall below our chosen
threshold. This is presumably due to differences in the model
potentials employed here and in \citet{ibata2021} to integrate orbits.

If we consider only the 60 stars in Table 1 with R.A.  $< 320\arcdeg$, scaling
the proper motions using the distances predicted by our model, and removing a
linear fit to the resulting velocity profile, we find an error-weighted,
two-dimensional velocity dispersion of $ 8.0 \pm 1.0$ km sec$^{-1}$. If we limit
the sample to only the 20 stars with the highest filter signal strengths, we
find $\sigma_{2d} = 3.8 \pm 0.6 $ km sec$^{-1}$. These values are somewhat
larger than the $\sim 2$ km sec$^{-1}$ 3-d velocity dispersions typical of other
cold streams and suggest that our sample as a whole suffers from at least some
contamination by field stars.  Radial velocity measurements will be needed to
remove these contaminants.

\section{Conclusions} \label{conclusions}

Using Gaia EDR3 photometry and proper motion measurements we have
detected a $45\arcdeg$-long stream of metal-poor stars that we believe
to be the trailing tidal tail of the globular cluster NGC 7089. The
detection was made possible by the fact that the proper motions
expected based on NGC 7089's orbit are significantly different from
those of most foreground stars in the vicinity.  The trajectory of the
trailing tail departs significantly from the integrated orbit of the
cluster, but is in good agreement with a model stream that has been
perturbed by the arrival of a massive LMC.

We cannot conclude from this that the departure of the trailing tail from its
predicted path in an unperturbed potential must be entirely due to the arrival
of the LMC. As the perigalacticon of the orbit lies less than 2 kpc from the
center of the Galaxy, the path of the stream could conceivably have been
modified by the bar \citep{price-whelan2016, sesar2016} or by interactions with
Giant Molecular Clouds \citep{amorisco2016}. The stream's current trajectory
could also be a consequence of tidal evolution of the cluster \citep{malhan2021}
while still orbiting its Gaia-Enceladus parent galaxy \citep{massari2019} prior
to its accretion by the Galaxy. Finally, we cannot rule out interactions with
other halo globular clusters, though such interactions would tend to produce
gaps or spurs rather than wholesale shifts in the trajectory.

The signal from the leading tidal tail is much less convincing, owing
to both its greater distance and to proper motions that are predicted
to be virtually identical to those of field stars.  By reducing the
proper motion and parallax uncertainties at fainter magnitudes, Gaia
DR3 and DR4 should be able to significantly reduce this degeneracy.

Verification of stream membership will require follow-up radial
velocity measurements. If even a few of the most outlying candidates
can be confirmed as having once belonged to NGC 7089, our modeling
suggests that this stream will become a particularly sensitive probe
of the shape of the halo potential in this Galactic quadrant, and a
significant contributor to our understanding of the influence of the
LMC and other components of the Galaxy.

\acknowledgments

We are very grateful to Rodgrigo Ibata for kindly providing the table
of positions and proper motions for the NGC 7089 stream stars detected
in \citet{ibata2021}. We are also grateful to an anonymous referee
whose careful reading of the manuscript significantly improved the
final product. This work has made use of data from the European Space
Agency (ESA) mission {\it Gaia}
(\url{https://www.cosmos.esa.int/gaia}), processed by the {\it Gaia}
Data Processing and Analysis Consortium (DPAC,
\url{https://www.cosmos.esa.int/web/gaia/dpac/consortium}). Funding
for the DPAC has been provided by national institutions, in particular
the institutions participating in the {\it Gaia} Multilateral
Agreement.

{\it Facilities:} \facility{Gaia, SDSS, LaMost}

\clearpage

\begin{figure}
\epsscale{1.0}
\plotone{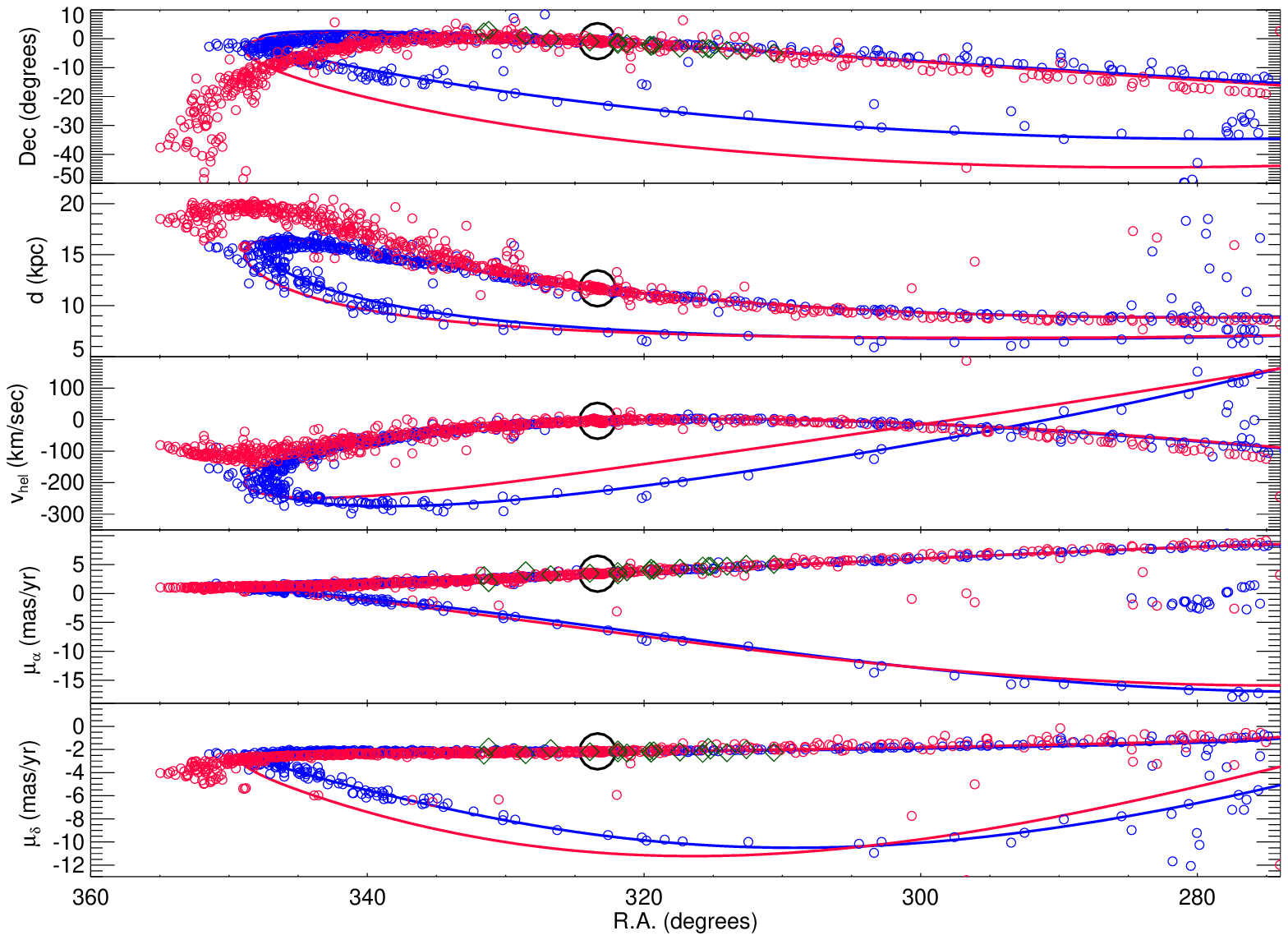}
\caption{The run of declination, distance, radial (line-of-sight) velocity, and
  proper motions with Right Ascension for the orbit of NGC 7089 and its tidal
  tails, as predicted using the Galactic model of \citet{allen1991}, updated
  with model parameters from \citet{irrgang2013}, and with or without a
  perturbing LMC. The curves show orbit integrations of NGC 7089 itself, using
  just the \citet{allen1991} model (blue curve) and supplemented by a point-mass
  LMC fixed at its current position with $ M = 1.88 \times 10^{11} M_\sun $ (red
  curve). The small open circles show the results of sequentially releasing 1000
  test particles from NGC 7089's L1 and L2 locations with a velocity dispersion
  of 1 km sec$^{-1}$ and integrating their orbits separately. The blue circles
  are test particles integrated using just the Galactic model, while the red
  circles include the effect of a massive LMC falling from 700 kpc to its
  current position on a first pass. The green diamonds show the positions and
  proper motions of stream stars detected by \citet{ibata2021} (kindly provided
  by R. Ibata). The large black circles show the measured quantities for NGC
  7089 itself. }

\end{figure}

\begin{figure}
\epsscale{1.0}
\plotone{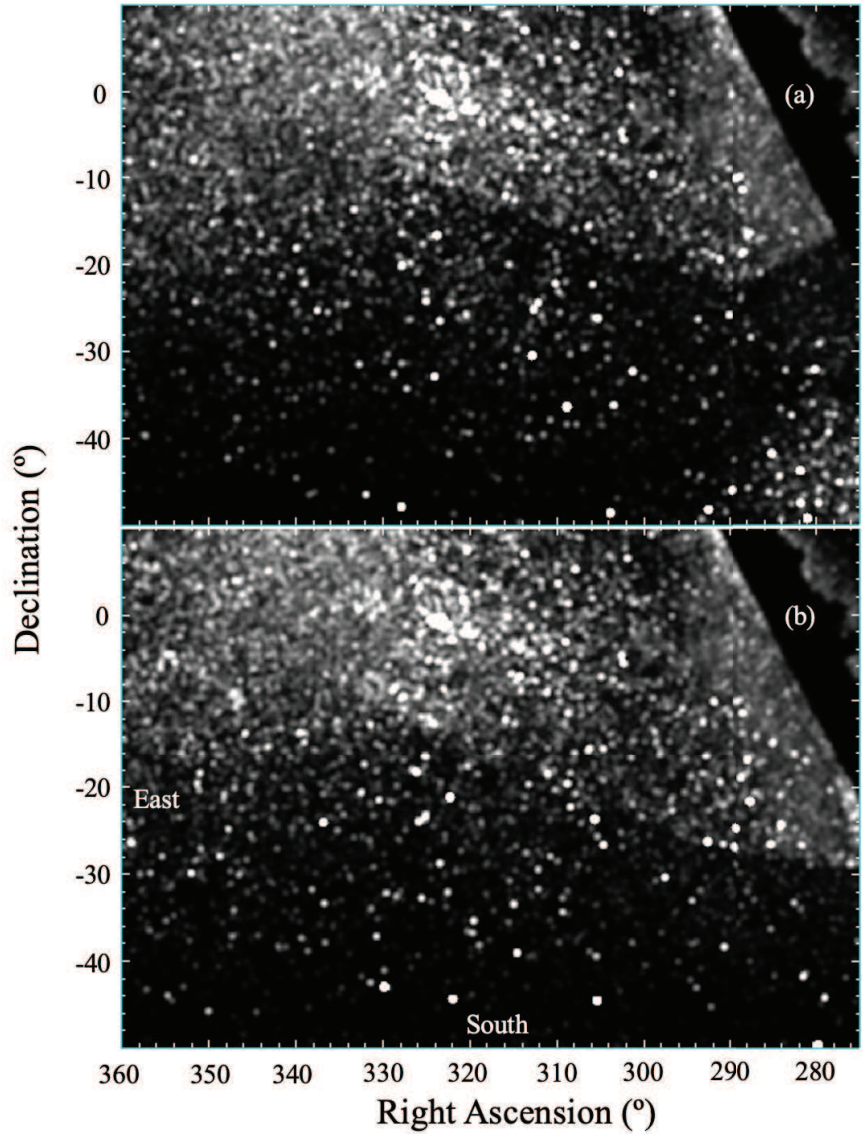}
\caption{Linear stretch of matched filter maps of the region
  surrounding NGC 7089. Individual stars are weighted by their
  positions in the NGC 7089 $G_\odot, (G_{BP} - G_{RP})_\odot$ color-magnitude
  diagram at a distance of 7.6 kpc (for $\alpha < 292\arcdeg$), 11.7
  kpc (for $292\arcdeg < \alpha < 334\arcdeg $) and 17.5 kpc (for
  $\alpha > 334\arcdeg$) and by their departure from the expected
  $\mu_\alpha$, $\mu_\delta$ profile of NGC 7089's tidal tails. The scale is
  $0.1\arcdeg$ per pixel and the maps have been smoothed with a Gaussian
  kernel of $0.3\arcdeg$. The background discontinuities arcing across the
  middle of the images are a consequence of applying different proper
  motion filters to areas near the leading and trailing tidal
  arms. The upper panel (a) shows the result of using the proper
  motion profiles in Figure 1 using just the \citet{allen1991} model
  of the Galactic potential, while the lower panel (b) shows the
  result of including a moving massive LMC.}
  \end{figure}

\begin{figure}
\epsscale{1.0}
\plotone{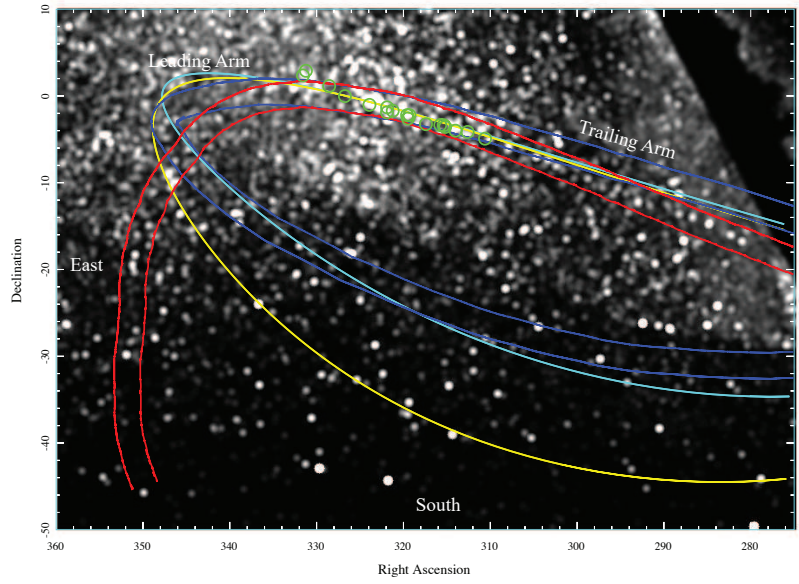}
\caption{Same as Panel (b) of Figure 2, but with computed trajectories
  superimposed. The cyan curve shows the predicted orbit of NGC 7089
  without taking account of the LMC, while the yellow curve shows the
  orbit in the presence of a massive LMC. The blue curves show
$\pm 1.5\arcdeg$ offsets from the mean trajectories of particles in our
stream models integrated without taking account of the LMC, while the red
curves indicate the mean orbital trajectories computed with a massive
LMC. The wobbles in this latter trajectory result from averaging
over the tidal feathers (formed by stars stripped during different
perigalactic passages), which can have slightly different paths on the
sky. The green circles show the NGC 7089 stream stars detected by 
\citet{ibata2021}. }
  \end{figure}

\begin{figure}
\epsscale{1.0}
\includegraphics[angle=90,width=7in]{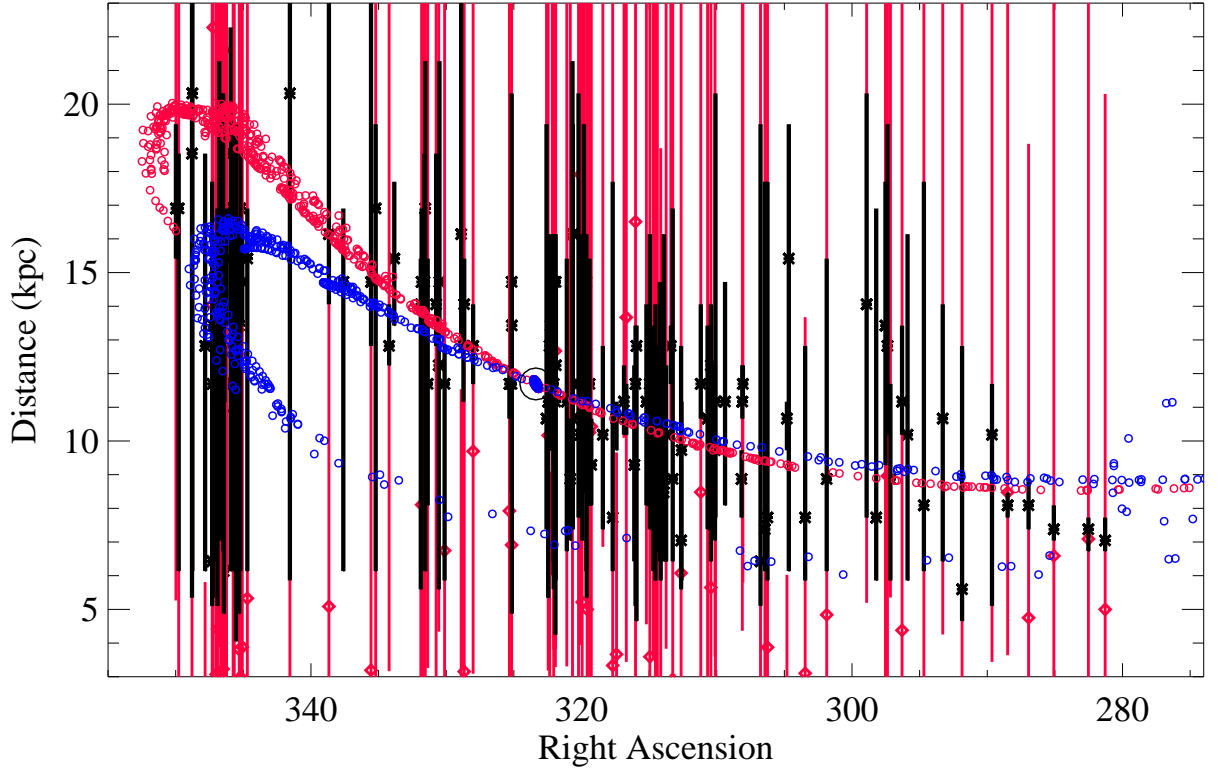}
\caption{Estimated distances of our highest-ranked candidate tail stars
  (black asterisks) compared with expectations from our unperturbed
  (blue circles) and moving LMC stream
  model (red circles). Black error bars denote $1\sigma$
  uncertainties based on the photometric match with NGC 7089's
  color-magnitude locus at the distance for which the matched-filter
  signal is at its maximum value. Red diamonds and red error bars show
  distances and the $1\sigma$
  uncertainties based on Gaia EDR3 parallax measurements. The open
  black circle shows NGC 7089 itself.}
\label{distance}
  \end{figure}

  \begin{figure}
\epsscale{1.0}
\includegraphics{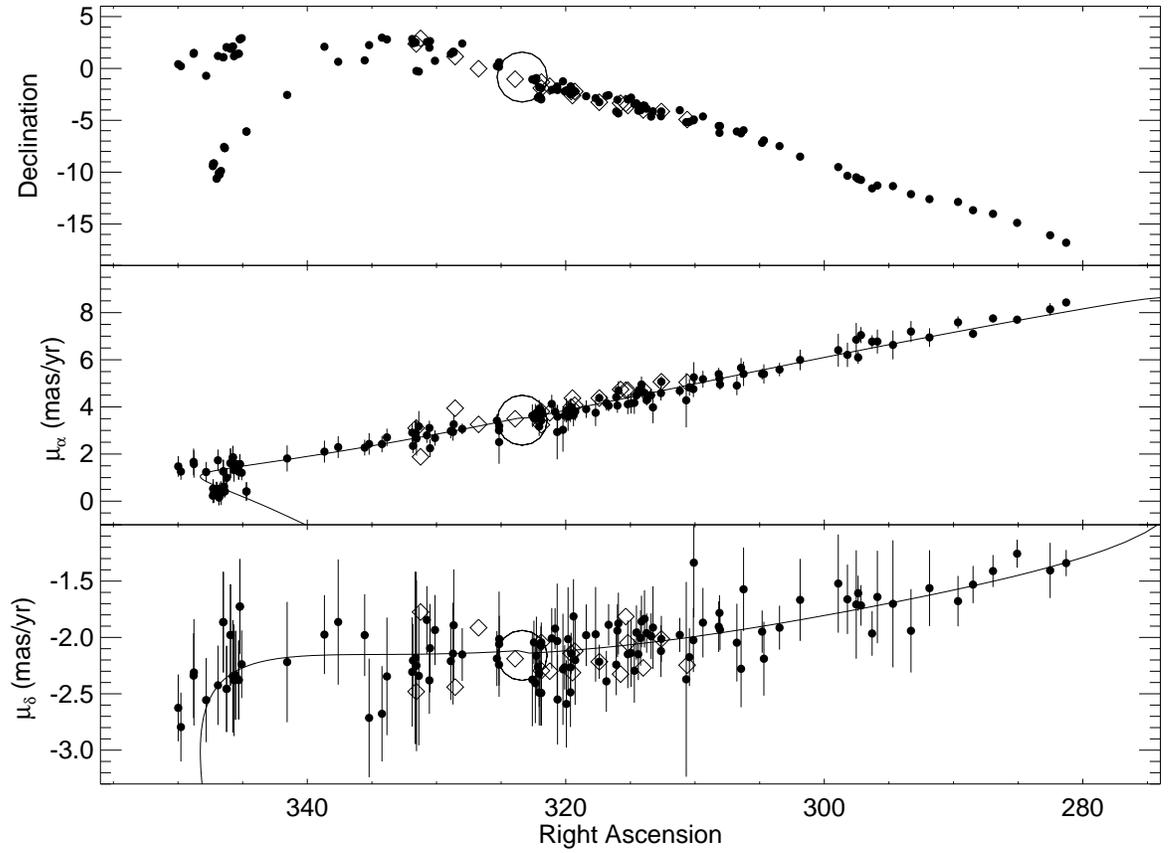}
\caption{Positions and proper motions of our highest-ranked candidate tail
  stars, as listed in Tables 1 and 2. The open circles denote
  NGC 7089 itself. The stars with R.A. $ > 344\arcdeg$ and dec $ > -2\arcdeg$ constitute the
  putative apogalactic ``cloud'', while the stars with R.A. $ >
  340\arcdeg$, dec $ < -2\arcdeg$ correspond to the leading tail
  predicted by the stream model in the presence of a massive LMC. The
  curves in the lower panels show the proper motion profiles used in
  our filter. Diamonds
  show the positions and proper motions of stream candidates detected by
  \citet{ibata2021}}
\label{positions}
  \end{figure}

\begin{deluxetable}{rccccccc}
\tabletypesize{\small}
\tablecaption{Candidate Stream Stars: Trailing Tail}
\tablecolumns{8}

\tablehead{ \colhead{No.} & \colhead{R.A. (J2016)} & \colhead{dec (J2016)} & \colhead{$G$} & \colhead{$G_{BP} - G_{RP}$} & \colhead{$\mu_\alpha \cos{\delta}$ (mas yr$^{-1}$)} & \colhead{$\mu_\delta$ (mas yr$^{-1}$)} & \colhead{Relative Weight}}\\
\startdata
\hline \\

  1 &   281.2847 &   -16.8027 &  17.912 &   1.149 &    8.431 $\pm$  0.141 &   -1.340 $\pm$  0.116 &  0.62 \\
  2 &   282.5242 &   -16.0834 &  18.799 &   1.269 &    8.141 $\pm$  0.265 &   -1.405 $\pm$  0.245 &  0.33 \\
  3 &   285.0792 &   -14.8886 &  18.075 &   1.126 &    7.701 $\pm$  0.149 &   -1.257 $\pm$  0.124 &  0.85 \\
  4 &   286.9467 &   -14.0203 &  18.071 &   1.110 &    7.754 $\pm$  0.166 &   -1.411 $\pm$  0.142 &  0.55 \\
  5 &   288.4917 &   -13.6707 &  18.251 &   0.943 &    7.103 $\pm$  0.185 &   -1.531 $\pm$  0.165 &  0.26 \\
  6 &   289.6521 &   -12.8711 &  19.106 &   0.832 &    7.588 $\pm$  0.258 &   -1.678 $\pm$  0.224 &  0.22 \\
  7 &   291.8608 &   -12.6001 &  19.334 &   0.923 &    6.946 $\pm$  0.394 &   -1.563 $\pm$  0.336 &  0.26 \\
  8 &   293.2917 &   -12.1231 &  19.608 &   0.835 &    7.197 $\pm$  0.448 &   -1.941 $\pm$  0.370 &  0.16 \\
  9 &   294.6858 &   -11.3475 &  19.894 &   0.875 &    6.630 $\pm$  0.613 &   -1.702 $\pm$  0.564 &  0.14 \\
 10 &   295.8914 &   -11.2883 &  19.781 &   0.858 &    6.773 $\pm$  0.509 &   -1.641 $\pm$  0.410 &  0.15 \\
 11 &   296.3004 &   -11.5659 &  18.683 &   1.004 &    6.769 $\pm$  0.270 &   -1.965 $\pm$  0.197 &  0.27 \\
 12 &   297.1597 &   -10.7432 &  19.245 &   0.830 &    7.046 $\pm$  0.340 &   -1.716 $\pm$  0.180 &  0.08 \\
 13 &   297.3797 &   -10.6547 &  18.688 &   1.106 &    6.096 $\pm$  0.265 &   -1.607 $\pm$  0.155 &  0.22 \\
 14 &   297.5283 &   -10.4981 &  20.264 &   0.835 &    6.851 $\pm$  0.712 &   -1.708 $\pm$  0.481 &  0.07 \\
 15 &   298.2094 &   -10.3495 &  19.810 &   0.892 &    6.209 $\pm$  0.516 &   -1.662 $\pm$  0.308 &  0.19 \\
 16 &   298.9180 &    -9.5097 &  20.168 &   0.791 &    6.405 $\pm$  0.696 &   -1.522 $\pm$  0.436 &  0.11 \\
 17 &   301.8697 &    -8.5042 &  19.593 &   0.774 &    5.993 $\pm$  0.436 &   -1.667 $\pm$  0.364 &  0.26 \\
 18 &   303.4628 &    -7.4861 &  19.156 &   0.710 &    5.585 $\pm$  0.284 &   -1.913 $\pm$  0.195 &  0.70 \\
 19 &   304.6718 &    -6.9360 &  19.739 &   0.694 &    5.397 $\pm$  0.408 &   -2.189 $\pm$  0.328 &  0.08 \\
 20 &   304.8095 &    -7.1719 &  18.757 &   0.732 &    5.399 $\pm$  0.233 &   -1.948 $\pm$  0.187 &  0.63 \\
 21 &   306.2461 &    -5.9585 &  19.610 &   0.745 &    5.401 $\pm$  0.517 &   -1.573 $\pm$  0.370 &  0.11 \\
 22 &   306.4382 &    -6.2603 &  19.612 &   0.748 &    5.659 $\pm$  0.419 &   -2.278 $\pm$  0.341 &  0.09 \\
 23 &   306.7650 &    -6.0792 &  19.593 &   0.784 &    4.906 $\pm$  0.409 &   -2.046 $\pm$  0.345 &  0.07 \\
 24 &   308.0669 &    -5.5551 &  18.824 &   0.775 &    4.953 $\pm$  0.229 &   -1.932 $\pm$  0.180 &  0.35 \\
 25 &   308.1076 &    -6.2084 &  18.615 &   0.823 &    5.202 $\pm$  0.201 &   -1.782 $\pm$  0.157 &  0.59 \\
 26 &   308.1644 &    -5.5455 &  18.881 &   0.634 &    5.391 $\pm$  0.266 &   -1.904 $\pm$  0.225 &  0.10 \\
 27 &   309.3865 &    -4.6300 &  19.494 &   0.665 &    5.176 $\pm$  0.356 &   -1.869 $\pm$  0.309 &  0.38 \\
 28 &   310.0923 &    -4.9250 &  19.977 &   0.705 &    5.254 $\pm$  0.642 &   -1.336 $\pm$  0.484 &  0.07 \\
 29 &   310.1293 &    -5.0035 &  19.134 &   0.663 &    4.757 $\pm$  0.344 &   -2.024 $\pm$  0.283 &  0.36 \\
 30 &   310.4335 &    -5.1539 &  19.277 &   0.697 &    4.829 $\pm$  0.355 &   -2.175 $\pm$  0.257 &  0.43 \\
 31 &   310.6783 &    -5.1780 &  19.346 &   0.676 &    4.281 $\pm$  1.153 &   -2.371 $\pm$  0.864 &  0.12 \\
 32 &   311.1795 &    -4.0251 &  18.419 &   0.886 &    4.679 $\pm$  0.223 &   -1.977 $\pm$  0.151 &  1.00 \\
 33 &   312.6071 &    -4.1474 &  18.892 &   0.708 &    5.068 $\pm$  0.236 &   -2.013 $\pm$  0.204 &  0.14 \\
 34 &   312.6337 &    -4.6105 &  19.180 &   0.757 &    4.585 $\pm$  0.314 &   -2.120 $\pm$  0.230 &  0.54 \\
 35 &   313.2539 &    -4.1308 &  19.649 &   0.741 &    3.980 $\pm$  0.673 &   -1.912 $\pm$  0.367 &  0.11 \\
 36 &   313.3965 &    -4.6425 &  19.129 &   0.809 &    4.497 $\pm$  0.269 &   -1.989 $\pm$  0.203 &  0.20 \\
 37 &   313.7426 &    -3.8802 &  19.080 &   0.707 &    4.285 $\pm$  0.226 &   -1.961 $\pm$  0.174 &  0.37 \\
 38 &   313.9071 &    -3.5868 &  19.548 &   0.737 &    4.594 $\pm$  0.367 &   -1.835 $\pm$  0.254 &  0.28 \\
 39 &   314.1409 &    -3.6345 &  19.335 &   0.753 &    4.948 $\pm$  0.340 &   -1.858 $\pm$  0.228 &  0.10 \\
 40 &   314.2102 &    -4.0325 &  19.376 &   0.726 &    4.765 $\pm$  0.309 &   -2.005 $\pm$  0.246 &  0.40 \\
 41 &   314.3891 &    -4.0681 &  19.225 &   0.692 &    4.652 $\pm$  0.272 &   -2.148 $\pm$  0.229 &  0.61 \\
 42 &   314.5338 &    -3.3642 &  19.290 &   0.731 &    4.489 $\pm$  0.377 &   -1.957 $\pm$  0.241 &  0.40 \\
 43 &   314.6923 &    -3.4327 &  19.205 &   0.723 &    4.172 $\pm$  0.469 &   -2.296 $\pm$  0.283 &  0.29 \\
 44 &   314.9455 &    -2.8081 &  19.606 &   0.699 &    4.155 $\pm$  0.429 &   -2.138 $\pm$  0.300 &  0.27 \\
 45 &   315.1995 &    -2.9921 &  19.527 &   0.681 &    4.111 $\pm$  0.421 &   -2.149 $\pm$  0.301 &  0.21 \\
 46 &   315.9309 &    -4.3343 &  19.090 &   0.762 &    4.697 $\pm$  0.395 &   -1.876 $\pm$  0.273 &  0.15 \\
 47 &   316.0043 &    -3.0211 &  19.178 &   0.732 &    4.057 $\pm$  0.309 &   -1.940 $\pm$  0.235 &  0.21 \\
 48 &   316.0946 &    -4.1570 &  19.108 &   0.686 &    4.413 $\pm$  0.292 &   -2.241 $\pm$  0.271 &  0.34 \\
 49 &   316.6855 &    -2.5764 &  18.721 &   0.789 &    4.052 $\pm$  0.241 &   -1.888 $\pm$  0.183 &  0.37 \\
 50 &   316.8570 &    -2.6468 &  18.836 &   0.775 &    4.162 $\pm$  0.320 &   -2.390 $\pm$  0.270 &  0.25 \\
 51 &   317.4079 &    -3.2508 &  18.484 &   0.835 &    4.377 $\pm$  0.177 &   -2.216 $\pm$  0.151 &  0.25 \\
 52 &   317.6899 &    -2.8665 &  19.596 &   0.733 &    3.750 $\pm$  0.564 &   -1.972 $\pm$  0.420 &  0.14 \\
 53 &   318.4211 &    -2.6732 &  19.248 &   0.669 &    3.903 $\pm$  0.374 &   -1.980 $\pm$  0.273 &  0.45 \\
 54 &   319.2730 &    -2.2189 &  19.041 &   0.650 &    3.801 $\pm$  0.385 &   -2.202 $\pm$  0.397 &  0.18 \\
 55 &   319.4027 &    -2.2788 &  19.561 &   0.619 &    3.637 $\pm$  0.458 &   -1.813 $\pm$  0.331 &  0.09 \\
 56 &   319.5807 &    -2.2141 &  18.679 &   0.731 &    3.993 $\pm$  0.227 &   -2.141 $\pm$  0.183 &  0.26 \\
 57 &   319.6214 &    -1.7309 &  19.019 &   0.707 &    3.827 $\pm$  0.330 &   -2.265 $\pm$  0.261 &  0.36 \\
 58 &   319.6348 &    -2.4786 &  19.512 &   0.726 &    3.840 $\pm$  0.480 &   -2.487 $\pm$  0.307 &  0.12 \\
 59 &   319.8227 &    -2.1205 &  20.002 &   0.632 &    3.583 $\pm$  0.598 &   -2.015 $\pm$  0.459 &  0.09 \\
 60 &   319.9501 &    -2.1123 &  19.579 &   0.677 &    3.853 $\pm$  0.513 &   -2.590 $\pm$  0.387 &  0.12 \\
 61 &   320.0846 &    -2.2095 &  18.668 &   0.735 &    3.635 $\pm$  0.265 &   -2.266 $\pm$  0.197 &  0.15 \\
 62 &   320.2185 &    -1.2443 &  19.993 &   0.697 &    3.029 $\pm$  0.929 &   -2.282 $\pm$  0.493 &  0.09 \\
 63 &   320.6388 &    -2.0746 &  19.899 &   0.695 &    3.587 $\pm$  0.480 &   -2.550 $\pm$  0.399 &  0.11 \\
 64 &   320.6637 &    -1.7388 &  20.046 &   0.717 &    2.934 $\pm$  1.159 &   -2.032 $\pm$  0.510 &  0.08 \\
 65 &   320.8262 &    -1.9968 &  18.914 &   0.661 &    3.790 $\pm$  0.268 &   -1.921 $\pm$  0.180 &  0.18 \\
 66 &   321.0859 &    -2.0591 &  19.428 &   0.678 &    4.126 $\pm$  0.390 &   -2.008 $\pm$  0.269 &  0.22 \\
 67 &   321.8929 &    -1.8965 &  18.653 &   0.886 &    3.805 $\pm$  0.276 &   -2.044 $\pm$  0.177 &  0.19 \\
 68 &   321.9056 &    -2.9722 &  19.478 &   0.745 &    3.428 $\pm$  0.343 &   -2.493 $\pm$  0.290 &  0.09 \\
 69 &   321.9276 &    -2.8502 &  18.777 &   0.758 &    3.674 $\pm$  0.238 &   -2.070 $\pm$  0.169 &  0.74 \\
 70 &   322.0541 &    -1.8172 &  19.554 &   0.674 &    3.169 $\pm$  0.405 &   -2.321 $\pm$  0.294 &  0.09 \\
 71 &   322.0563 &    -2.6697 &  19.437 &   0.744 &    3.921 $\pm$  0.376 &   -2.490 $\pm$  0.291 &  0.10 \\
 72 &   322.1595 &    -2.8156 &  18.994 &   0.774 &    3.761 $\pm$  0.270 &   -2.261 $\pm$  0.231 &  0.21 \\
 73 &   322.2721 &    -0.9314 &  19.851 &   0.583 &    3.556 $\pm$  0.449 &   -2.163 $\pm$  0.348 &  0.11 \\
 74 &   322.3511 &    -1.1634 &  19.781 &   0.632 &    3.700 $\pm$  0.424 &   -2.406 $\pm$  0.354 &  0.13 \\
 75 &   322.4616 &    -1.0223 &  19.018 &   0.706 &    3.526 $\pm$  0.233 &   -2.044 $\pm$  0.193 &  0.75 \\
 76 &   322.5867 &    -1.0529 &  19.899 &   0.705 &    3.627 $\pm$  0.479 &   -2.374 $\pm$  0.413 &  0.19 \\

\enddata
\end{deluxetable}

\begin{deluxetable}{rccccccc}
\tabletypesize{\small}
\tablecaption{Candidate Stream Stars: Leading Tail}
\tablecolumns{8}

\tablehead{ \colhead{No.} & \colhead{R.A. (J2016)} & \colhead{dec (J2016)} & \colhead{$G$} & \colhead{$G_{BP} - G_{RP}$} & \colhead{$\mu_\alpha \cos{\delta}$ (mas yr$^{-1}$)} & \colhead{$\mu_\delta$ (mas yr$^{-1}$)} & \colhead{Relative Weight}}\\
\startdata
\hline \\

 77 &   325.1483 &     0.5728 &  19.222 &   0.778 &    3.193 $\pm$  0.361 &   -2.014 $\pm$  0.205 &  0.21 \\
 78 &   325.1571 &     0.5836 &  20.055 &   0.685 &    2.507 $\pm$  0.921 &   -2.059 $\pm$  0.467 &  0.07 \\
 79 &   325.1670 &     0.1508 &  18.549 &   0.882 &    3.006 $\pm$  0.191 &   -2.241 $\pm$  0.185 &  0.20 \\
 80 &   325.3284 &     0.2358 &  18.827 &   0.773 &    3.420 $\pm$  0.265 &   -2.187 $\pm$  0.244 &  0.64 \\
 81 &   328.0100 &     2.4013 &  18.878 &   0.854 &    3.064 $\pm$  0.227 &   -2.151 $\pm$  0.234 &  0.40 \\
 82 &   328.6698 &     1.5576 &  19.161 &   0.808 &    3.262 $\pm$  0.404 &   -1.892 $\pm$  0.495 &  0.19 \\
 83 &   328.7255 &     1.6134 &  19.071 &   0.784 &    2.936 $\pm$  0.367 &   -2.144 $\pm$  0.451 &  0.18 \\
 84 &   328.8938 &     1.3983 &  18.886 &   0.911 &    2.976 $\pm$  0.306 &   -2.210 $\pm$  0.345 &  0.38 \\
 85 &   330.1190 &     0.7344 &  19.107 &   0.704 &    2.682 $\pm$  0.329 &   -1.934 $\pm$  0.311 &  0.43 \\
 86 &   330.5000 &     2.6337 &  19.183 &   0.843 &    2.244 $\pm$  0.374 &   -2.095 $\pm$  0.392 &  0.09 \\
 87 &   330.5507 &     2.0030 &  19.141 &   0.705 &    3.105 $\pm$  0.306 &   -2.381 $\pm$  0.296 &  0.21 \\
 88 &   330.7462 &     2.5520 &  19.002 &   0.853 &    2.795 $\pm$  0.322 &   -1.845 $\pm$  0.300 &  0.22 \\
 89 &   331.3680 &    -0.2976 &  19.662 &   0.671 &    3.172 $\pm$  0.651 &   -2.341 $\pm$  0.616 &  0.08 \\
 90 &   331.5506 &    -0.2414 &  19.984 &   0.789 &    2.668 $\pm$  0.733 &   -2.250 $\pm$  0.759 &  0.07 \\
 91 &   331.6195 &     2.5053 &  19.252 &   0.824 &    2.885 $\pm$  0.516 &   -2.181 $\pm$  0.764 &  0.14 \\
 92 &   331.8195 &     2.4248 &  19.223 &   0.821 &    2.344 $\pm$  0.307 &   -2.205 $\pm$  0.324 &  0.26 \\
 93 &   331.8793 &     2.8398 &  19.334 &   0.706 &    2.909 $\pm$  0.403 &   -2.306 $\pm$  0.485 &  0.14 \\
 94 &   333.8377 &     2.8016 &  19.324 &   0.820 &    2.706 $\pm$  0.373 &   -2.346 $\pm$  0.522 &  0.12 \\
 95 &   334.2209 &     2.9675 &  19.049 &   0.814 &    2.423 $\pm$  0.349 &   -2.677 $\pm$  0.423 &  0.09 \\
 96 &   335.2148 &     2.2480 &  19.816 &   0.755 &    2.428 $\pm$  0.459 &   -2.714 $\pm$  0.526 &  0.09 \\
 97 &   335.5576 &     0.7801 &  19.042 &   0.908 &    2.272 $\pm$  0.335 &   -1.979 $\pm$  0.363 &  0.26 \\
 98 &   337.6078 &     0.6448 &  19.612 &   0.724 &    2.294 $\pm$  0.464 &   -1.863 $\pm$  0.555 &  0.12 \\
 99 &   338.6724 &     2.0934 &  19.404 &   0.874 &    2.102 $\pm$  0.464 &   -1.975 $\pm$  0.350 &  0.17 \\
100 &   341.5561 &    -2.5554 &  20.008 &   0.776 &    1.812 $\pm$  0.553 &   -2.218 $\pm$  0.534 &  0.08 \\
101 &   344.7044 &    -6.0849 &  19.533 &   0.733 &    0.412 $\pm$  0.395 &   -4.482 $\pm$  0.335 &  0.22 \\
102 &   345.0743 &     2.9189 &  19.265 &   0.815 &    1.204 $\pm$  0.316 &   -2.238 $\pm$  0.300 &  0.18 \\
103 &   345.2132 &     2.8187 &  19.728 &   0.737 &    1.573 $\pm$  0.419 &   -1.726 $\pm$  0.421 &  0.08 \\
104 &   345.3032 &     1.4221 &  19.359 &   0.713 &    1.258 $\pm$  0.350 &   -2.377 $\pm$  0.350 &  0.08 \\
105 &   345.5452 &     1.3684 &  19.188 &   0.812 &    1.569 $\pm$  0.343 &   -2.336 $\pm$  0.394 &  0.14 \\
106 &   345.6678 &     1.1773 &  19.850 &   0.685 &    1.320 $\pm$  0.493 &   -2.380 $\pm$  0.496 &  0.13 \\
107 &   345.7548 &     2.1181 &  19.748 &   0.756 &    1.859 $\pm$  0.486 &   -2.345 $\pm$  0.498 &  0.11 \\
108 &   345.9409 &     1.9106 &  19.944 &   0.751 &    1.614 $\pm$  0.623 &   -1.978 $\pm$  0.450 &  0.08 \\
109 &   346.2401 &     2.0400 &  19.441 &   0.786 &    0.995 $\pm$  0.392 &   -2.456 $\pm$  0.382 &  0.10 \\
110 &   346.3858 &    -7.6764 &  18.854 &   0.779 &    0.415 $\pm$  0.254 &   -4.253 $\pm$  0.239 &  0.26 \\
111 &   346.4380 &    -7.5666 &  19.149 &   0.741 &    0.624 $\pm$  0.369 &   -4.400 $\pm$  0.269 &  0.32 \\
112 &   346.4946 &     1.0766 &  19.903 &   0.743 &    1.261 $\pm$  0.498 &   -1.864 $\pm$  0.446 &  0.08 \\
113 &   346.6741 &    -9.8808 &  19.894 &   0.653 &    0.340 $\pm$  0.506 &   -4.534 $\pm$  0.453 &  0.09 \\
114 &   346.7813 &   -10.2359 &  19.066 &   0.860 &    0.530 $\pm$  0.260 &   -4.757 $\pm$  0.288 &  0.26 \\
115 &   346.8299 &   -10.0795 &  19.216 &   0.800 &    0.160 $\pm$  0.334 &   -4.817 $\pm$  0.337 &  0.30 \\
116 &   346.9086 &     1.2022 &  19.415 &   0.762 &    1.734 $\pm$  0.458 &   -2.424 $\pm$  0.352 &  0.12 \\
117 &   347.0098 &   -10.6225 &  19.197 &   0.669 &    0.476 $\pm$  0.346 &   -4.916 $\pm$  0.275 &  0.09 \\
118 &   347.2283 &    -9.1633 &  19.249 &   0.710 &    0.293 $\pm$  0.356 &   -4.412 $\pm$  0.397 &  0.22 \\
119 &   347.2848 &    -9.1872 &  19.562 &   0.634 &    0.528 $\pm$  0.407 &   -4.568 $\pm$  0.407 &  0.15 \\
120 &   347.3029 &    -9.4175 &  19.481 &   0.763 &    0.232 $\pm$  0.316 &   -4.241 $\pm$  0.317 &  0.10 \\
121 &   347.8227 &    -0.7089 &  19.680 &   0.657 &    1.236 $\pm$  0.430 &   -2.556 $\pm$  0.373 &  0.09 \\
122 &   348.7701 &     1.5012 &  20.005 &   0.751 &    1.573 $\pm$  0.576 &   -2.310 $\pm$  0.472 &  0.07 \\
123 &   348.7943 &     1.4163 &  19.738 &   0.792 &    1.669 $\pm$  0.566 &   -2.339 $\pm$  0.376 &  0.07 \\
124 &   349.7764 &     0.2186 &  19.669 &   0.760 &    1.255 $\pm$  0.350 &   -2.795 $\pm$  0.305 &  0.19 \\
125 &   349.9826 &     0.4109 &  19.531 &   0.812 &    1.475 $\pm$  0.436 &   -2.625 $\pm$  0.296 &  0.15 \\

\enddata
\end{deluxetable}


\begin{thebibliography}{}

\bibitem[Aguado et al.(2018)]{aguado2018} Aguado et al. 2018, \apjs,
  240, 23

\bibitem[Allen \& Santillan(1991)]{allen1991} Allen, C., \& Santillan,
  A. 1991, {\it Rev. Mex. Astron. Astrofis.}, 22, 255

\bibitem[Amorisco et al.(2016)]{amorisco2016} Amorisco, N. C. et
al. 2016, \mnras, 463, 17

\bibitem[Balbinot \& Gieles(2018)]{balbinot2017} Balbinot, E. \&
  Gieles, M. 2018, \mnras, 474, 2479

\bibitem[Baumgardt \& Vasiliev(2021)]{baumgardt2021} Baumgardt, H. \&
  Vasiliev, E. 2021, \mnras, 505, 5957

\bibitem[Bonaca(2018)]{bonaca2018} Bonaca, A., \& Hogg, D. W. 2018,
  \apj, 867, 101

\bibitem[Bovy et al.(2016)]{bovy2016} Bovy, J., Bahmanyar, A., Fritz,
  T. K., \& Kallivayalil, N. 2016, \apj, 833, 31

\bibitem[Bowden et al.(2015)]{bowden2015} Bowden, A., Belokurov, V.,
  \& Evans, N. W. 2015, /mnras, 449, 1391

 \bibitem[Chandrasekhar(1943)]{chandrasekhar1943} Chandrasekhar,
   S. 1943, \apj, 98, 54 

 \bibitem[Conroy et al.(2021)]{conroy2021} Conroy, C. et al. 2021,
   Nature, 592, 534

 \bibitem[Dai et al.(2018)]{dai2018} Dai, B., Robertson, B. E., \&
   Madau, P. 2018, \apj, 858, 73

\bibitem[Erkal et al.(2019)]{erkal2019} Erkal, D., Belokurov, C.,
  Laport, C. F. P. et al. 2019, \mnras, 487, 2685

\bibitem[Gaia Collaboration et al.(2021)]{gaia2021} Gaia
  Collaboration, Brown, A. et al. 2021, \aap, 649, 1

\bibitem[Gaia Collaboration et al.(2018)]{gaia2018} Gaia
  Collaboration, Babusiaux, C., van Leeuwen, F. et al. 2018, \aap, 616, A10

\bibitem[Garavito-Camargo et al.(2021)]{garavito-camargo2021}
  Garavito-Camargo, N. et al. 2021, \apj, 919, 109

\bibitem[Gialluca et al.(2021)]{gialluca2021} Gialluca, M. T., Naidu,
  R. P., \& Bonaca, A. 2021, \apjl, 911, 32  

 \bibitem[Grillmair \& Dionatos(2006)]{grillmair2006} Grillmair,
   C. J. \& Dionatos, O. 2006, \apjl, 639,  17

\bibitem[Grillmair(2009)]{grillmair2009} Grillmair, C. J. 2009, \apj,
  694, 1118

\bibitem[Grillmair(2019)]{grillmair2019} Grillmair, C. J. 2019, \apj,
  884, 174

\bibitem[Grillmair et al.(1995)]{grillmair1995} Grillmair, C. J.,
  Freeman, K. C., Irwin, M., \& Quinn, P. J. 1995, \aj, 109, 2553

\bibitem[Grillmair \& Carlin(2016)]{grillmair2016} Grillmair, C. J.,
  \& Carlin, J. L. 2016, in {\it Tidal Streams in the Local Group and
    Beyond}, H. J. Newberg \& J. L. Carlin eds., Springer

\bibitem[Harris(1996)]{harris1996} Harris, W. E., 1996, \aj, 112, 1487  

\bibitem[Ibata et al.(2019)]{ibata2019} Ibata, R. A., Malhan, K., \&
  Martin, N. F. 2019, \apj, 872, 152

  \bibitem[Ibata et al.(2021)]{ibata2021} Ibata, R. A. et al. 2021,
    \apj, 914, 1231

\bibitem[Irrgang et al.(2013)]{irrgang2013} Irrgang, A., Wilcox, B.,
  Tucker, E., \& Schiefelbein, L. 2013, \aa, 549, 137

\bibitem[Kallivayalil et al.(2013)]{kallivayalil2013} Kallivayalil,
  N. et al. 2013, \apj, 764, 161  

\bibitem[K{\"u}pper et al.(2012)]{kupper2012} K{\"u}pper, A. H. W., Lane,
  R. R., \& Heggie, D. C. 2012, \mnras, 420, 2700


\bibitem[K{\"u}pper et al.(2015)]{kupper2015} K{\"u}pper, A. H. W.,
    et al. 2015, \apj, 803, 80

  \bibitem[Kuzma et al.(2016))]{kuzma2016} Kuzma, P. B., Da Costa, G. S., Mackey, A. D., \& Roderick, T. A. 2016, \mnras, 461, 3639

\bibitem[Malhan \& Ibata(2018)]{malhan2018a} Malhan, K. \& Ibata,
  R. A. 2018, \mnras, 477, 4063

\bibitem[Malhan et al.(2018)]{malhan2018b} Malhan, K., Ibata, R. A.,
  \& Martin, N. F. 2018, \mnras, 481, 3442

\bibitem[Malhan \& Ibata(2019)]{malhan2019} Malhan, K. \& Ibata,
  R. A., 2019, \mnras, 486, 2995

\bibitem[Malhan et al.(2019)]{malhan2019a}   Malhan, K., et al. 2019,
  \apj, 881, 106

\bibitem[Malhan et al.(2021)]{malhan2021}  Malhan, K., Valluri, M., \&
  Freese, K. 2021, \mnras, 501, 179

\bibitem[Massari et al.(2019)]{massari2019} Massari, D., Koppelman,
  H. H., \& Helmi, A. 2019, \aap, 630, 4

\bibitem[Miyamoto \& Nagai(1975)]{miyamoto1975} Miyamoto, M. \& Nagai,
  R., 1975, Pub.Astr.Soc.Japan, 27, 533

 \bibitem[Navarro et al.(1997)]{navarro1997} Navarro, J. F., Frenk,
   C. S., \& White, S. D. M., 1997, \apj, 490, 493

\bibitem[Odenkirchen et al.(2001)]{odenkirchen2001} Odenkirchen, M. et
  al. 2001, \apjl, 548, 165

\bibitem[Odenkirchen et al.(2009)]{odenkirchen2009} Odenkirchen, M. et
  al. 2009, \aj, 137, 3378

\bibitem[Palau \& Miralda-Escude(2019)]{palau2019} Palau, C. G. \&
  Miralda-Escude, J. 2019, arXiv: 1905.01193

\bibitem[Price-Whelan et al.(2014)]{price-whelan2014} Price-Whelan,
  A. M. et al. 2014, \apj, 794, 4

\bibitem[Price-Whelan et al.(2016)]{price-whelan2016} Price-Whelan,
  A. M. et al. 2016, \apj, 824, 104

  
\bibitem[Rockosi et al.(2002)]{rockosi2002} Rockosi, C. M.,
  Odenkirchen, M., Grebel, E. K. et al. 2002, \aj, 124, 349

\bibitem[Schlafly \& Finkbeiner(2011)]{schlafly2011} Schlafly, E. F.,
  \& Finkbeiner, D. P. 2011, \apj, 737, 103

\bibitem[Schlegel et al.(1998)]{schlegel1998} Schlegel, D. J., Finkbeiner, D. P., \& Davis, M. 1998, \apj, 500, 525

 \bibitem[Sesar et al.(2016)]{sesar2016} Sesar, B. et al. 2016, \apj,
   816, 4 

\bibitem[Shipp et al.(2018)]{shipp2018} Shipp, N., Drlica-Wagner, A.,
  Balbinot, E. et al. 2018, \apj 862, 114

\bibitem[Shipp et al.(2021)]{shipp2021} Shipp, N. Erkal, D.,
  Drlica-Wagner, A. et al. 2021, \apj, 923, 149

\bibitem[Vasiliev(2019)]{vasiliev2019} Vasiliev, E. 2019, \mnras, 484, 2832  
  
\bibitem[Vasiliev \& Baumgardt(2021)]{vasiliev2021} Vasiliev, E., \&
  Baumgardt. H., 2021, \mnras, 505, 5978

\bibitem[Wilkinson  \& Evans(1999)]{wilkinson1999} Wilkinson, M. I.,
    \& Evans, N. W., 1999, \mnras, 310, 645

\bibitem[Zhao et al.(2012)]{zhao2012} Zhao, G. et al. 2012, arXiv:1206.3569
\end{thebibliography}
\end{document}